\documentclass[%
preprint,
nofootinbib,
 amsmath,amssymb,
 aps,
]{revtex4-2}

\usepackage{graphicx}
\usepackage{dcolumn}
\usepackage{bm}

\begin{document}

\def\bb    #1{\hbox{\boldmath${#1}$}}

\title{Scaling for count-in-cell and factorial moment analysis}

\author{Valeria Zelina Reyna Ortiz} 
\email{val.reyna@cern.ch}
\affiliation{Institute of Physics, Jan Kochanowski University, 25-406 Kielce, Poland}
\author{Maciej Rybczy\'nski}
\email{maciej.rybczynski@ujk.edu.pl}
\affiliation{Institute of Physics, Jan Kochanowski University, 25-406 Kielce, Poland}
\author{Zbigniew W\l odarczyk}
\email{zbigniew.wlodarczyk@ujk.edu.pl}
\affiliation{Institute of Physics, Jan Kochanowski University, 25-406 Kielce, Poland}

\begin{abstract} 
The investigation of remnants associated with the QCD chiral critical point is a primary objective in high-energy ion collision experiments. Numerous studies indicate that a scaling relation between higher-order factorial moments of hadron multiplicity distributions and the second factorial moment may serve as a diagnostic tool for identifying the QCD critical point. However, we demonstrate that this scaling behavior is not exclusive to critical phenomena but rather arises as a general consequence of the phase-space partitioning procedure employed in the analysis. This finding is examined in the context of recent intermittency analyses conducted by the STAR experiment at RHIC.
\end{abstract}

\pacs{XXX}

\maketitle

\section{Introduction}
\label{Introduction}

The primary objective of intermittency analysis, which investigates self-similar correlations as a function of phase-space volume size, in relativistic heavy-ion collision programs is to probe the phase diagram of quantum chromodynamic (QCD) matter. A fundamental feature of this phase diagram is the critical endpoint (CEP), which marks the termination of the first-order phase transition boundary~\cite{Bowman:2008kc, Hatta:2003wn}. Extensive efforts have been undertaken to identify the possible location of the CEP in heavy-ion collision experiments~\cite{Bzdak:2019pkr, Luo}.

Recent discussions have considered the potential role of particle density fluctuations in heavy-ion collisions as a distinctive signature of the phase transition within the QCD phase diagram~\cite{Li:2015pbv, Sun:2018jhg}. Analogous to the phenomenon of critical opalescence, it is expected that the produced medium will exhibit significant particle density fluctuations near the CEP due to the rapid growth of the correlation length within the critical region~\cite{Stephanov:1998dy, Berdnikov:1999ph}. If these large-scale density fluctuations persist through the kinetic freeze-out stage and remain unaffected by final-state interactions during the hadronic evolution of the system, they may serve as an observable signal of critical behavior.
Upon approaching a critical point, the correlation length of the system diverges, leading to scale invariance and self-similarity~\cite{DeWolf:1995nyp, Bialas:1990xd, Satz:1989vj}. Within the framework of the 3D-Ising universality class,
 the density-density correlation function for small momentum transfer follows a power-law behavior, giving rise to significant density fluctuations in heavy-ion collisions~\cite{Antoniou:2006zb, Antoniou:2017vti, Antoniou:2000ms, Antoniou:2015lwa}. These fluctuations can be investigated through an intermittency analysis in transverse momentum phase space using scaled factorial moments~\cite{Antoniou:2006zb, Antoniou:2017vti, Antoniou:2000ms, Antoniou:2015lwa, NA49:2009diu, Wu:2019mqq}.

The methodology involves dividing the phase space into $M$ equal-sized cells, where the $q$-th order scaled factorial moment, $F_{q}\left(M\right)$, is defined as~\cite{Antoniou:2006zb, Hwa:1992uq, NA49:2012ebu, Antoniou:2005am, Bialas:1985jb, Bialas:1988wc}:
\begin{equation} 
F_{q}\left(M\right) = \frac{\langle \frac{1}{M}\sum_{i=1}^{M} n_{i}\left(n_{i}-1\right)\ldots\left(n_{i}-q+1\right) \rangle}{\langle \frac{1}{M} \sum_{i=1}^{M} n_{i} \rangle^{q}}, 
\label{Fq} 
\end{equation}
where $M$ represents the number of phase-space cells, and $n_{i}$ denotes the measured multiplicity of a given event in the $i$-th cell. The angle brackets indicate an average over all events.
To mitigate background contributions, including fluctuations in overall multiplicity, the "mixed event method" is commonly employed~\cite{NA49:2009diu, NA49:2012ebu, Antoniou:2005am, STAR:2023jpm, Wu:2022aio}. In this approach, mixed events are generated by randomly selecting particles from different original events while maintaining the same multiplicity and momentum distributions as the original events. Consequently, rather than using $F_q(M)$ directly, the following observable is typically analyzed:
\begin{equation} \Delta F_{q}\left(M\right) =F_{q}\left(M\right)^{data}-F_{q}\left(M\right)^{mix}, 
\label{deltaFq} 
\end{equation}
where the factorial moments obtained from mixed events are subtracted from those measured in actual data.

Intermittency manifests as a power-law scaling of scaled factorial moments, with factorial moments increasing as the phase-space cell size decreases~\cite{Antoniou:2006zb, Hwa:1992uq, Antoniou:2000ms, Bialas:1985jb}. In systems exhibiting density fluctuations, the scaled factorial moments follow the relation:
\begin{equation} 
F_{q}\left(M\right) \propto \left(M \right)^{\phi_{q}}, \quad M\gg 1, 
\end{equation}
where $\phi_{q}$ is the intermittency index, quantifying the strength of intermittency~\cite{Antoniou:2006zb, Antoniou:2017vti, Antoniou:2000ms, NA49:2009diu, NA49:2012ebu}.

In the context of particle physics, factorial moment analysis is particularly sensitive to critical fluctuations, exhibiting intermittency~\cite{Satz:1989vj}, a phenomenon analogous to critical opalescence in strongly interacting matter~\cite{Antoniou:2006zb}. Within this framework, intermittency is characterized by the scaling law (valid for sufficiently large $M$):
\begin{equation} 
\Delta F_q(M) \propto M^{(q-1)d_F}, 
\label{Fq1} 
\end{equation}
where $d_F$ is the fractal dimension associated with the geometric structure of particle clusters in momentum space. However, achieving a pure monofractal behavior in factorial moment analysis of particle momenta from ion collisions is challenging due to finite statistics, which prevent reaching the theoretical limit of $M\rightarrow \infty$, where the intermittency effect is expected to be fully realized.

As an alternative method for detecting the QCD critical endpoint through intermittency analysis, it was proposed in~\cite{Hwa:1992uq} to investigate the scaling of the ratio of factorial moments. The scaling law in Eq.~(\ref{Fq1}) implies a related scaling relation between the $q$-th factorial moment and the second factorial moment:
\begin{equation} 
\Delta F_q(M) \propto \Delta F_2(M)^{(q-1)}.
\label{Fq2} 
\end{equation}
Thus, if intermittency is present in accordance with Eq.~(\ref{Fq1}), and the detected particle momenta originate from a source at a critical state, Eq.~(\ref{Fq2}) is expected to hold. Furthermore, the fractal dimension $d_F$ in Eq.~(\ref{Fq1}) is determined by the isothermal critical exponent associated with the universality class of the transition~\cite{Antonou:JPG}.

Using a Ginzburg-Landau free energy formulation to describe order parameter density fluctuations, it has been demonstrated that a more general scaling relation holds:
\begin{equation} 
\Delta F_q(M) \propto \Delta F_2(M)^{\beta_q},
\label{Fq3} 
\end{equation}
where $\beta_q = (q-1)^{\nu}$ with $ \nu \approx 1.304$. This scaling law is particularly valid in the symmetry-broken (hadronic) phase. Based on this result, it was proposed in~\cite{Hwa:1992uq} to search for the scaling behavior in Eq.~(\ref{Fq3}) as an indicator of proximity to the critical point from the hadronic phase. The key objective in this approach is to measure the exponent $\nu$, with the expectation that it should be close to 1.3. This behavior contrasts with Eq.~(\ref{Fq2}), which suggests $\nu = 1$ under the assumption that the particle source at the critical point exhibits a monofractal structure.

Recently, this alternative method was employed in an analysis by the STAR Collaboration at RHIC, using charged-particle momenta from Au+Au collisions across different beam energies and centralities~\cite{STAR:2023jpm}. Surprisingly, the analysis revealed a scaling behavior consistent with Eq.~(\ref{Fq3}) across all examined systems, but with a $\nu$-value significantly smaller than 1, exhibiting slight variations with beam energy and centrality. The origin of this unexpected behavior was not addressed in~\cite{STAR:2023jpm}, raising questions regarding its physical interpretation.

The use of factorial moments of hadron multiplicity fluctuations as a diagnostic tool for identifying the QCD critical point remains a subject of debate. Recent studies have demonstrated that the observed scaling law for factorial moments arises as a general property of multiplicity distributions, utilizing the concept of hypercontractivity applied to factorial moments~\cite{Brofas}.

In~\cite{NPA}, we demonstrated that factorial moment analysis, as a function of the phase-space cell size, exhibits sensitivity to particle clustering when the cell size is reduced but remains sufficiently large. Specifically, density fluctuations within individual cells give rise to a power-law-like behavior in scaled factorial moments. A correlated cluster of particles confined to a single phase-space cell can induce an "intermittency-like" pattern in $\Delta F_{q}\left(M\right)$.

Based on this scenario, we further examine the phase-space division procedure and argue that the count-in-cell approach inherently leads to a power-law dependence of $\Delta F_{q}\left(M\right)$. This theoretical insight is then compared with recent intermittency analysis results from the STAR experiment at RHIC~\cite{STAR:2023jpm}.




\section{Particles in cells}
\label{one bunch}

For a large number of phase-space cells, $M\gg N$ (where $N=\sum n_{i}$), the average multiplicity per cell, $\langle n_{i}\rangle$, becomes small, typically resulting in either zero or one particle 
being observed in a given cell. Under these conditions, the multiplicity distributions within individual cells become indistinguishable. Considering two limiting cases of multiplicity distributions, the 
Poisson distribution (PD) and the geometric (Bose-Einstein (BE)) distribution, one finds that for small $\langle n_{i}\rangle$, the probabilities of observing zero or one particle are given by
$P(0) = \exp(-\langle n_{i}\rangle) \simeq 1 - \langle n_{i}\rangle$, $P(1) = \langle n_{i}\rangle \exp(-\langle n_{i}\rangle) \simeq \langle n_{i}\rangle$ for the Poisson distribution. 
Similarly, for the geometric distribution: $P(0) = \frac{1}{1+\langle n_{i}\rangle} \simeq 1 - \langle n_{i}\rangle$, $P(1) = \frac{\langle n_{i}\rangle}{(1+\langle n_{i}\rangle)^{2}} 
\simeq \langle n_{i}\rangle$.
In contrast, the probability of detecting a cluster of particles with $n > 1$ is enhanced by a factor of $n!$ in the geometric distribution compared to the Poisson distribution. This relationship can be expressed as:
$P_{BE}(n_i) \simeq n_i! (1+\langle n_i \rangle)^{-n_i} P_{PD}(n_i).$

Assuming that a  cluster of correlated  particles appears within a single cell, the factorial moment can be approximated (for a large number of cells) as $F_{q} \simeq F_{q}^{PD}+\langle f_q^{}\rangle\left(\langle n_{i}\rangle^{q} M\right)^{-1}$. 
From this, the difference in factorial moments between data and mixed events is given by: 
\begin{equation} 
\Delta F_{q} \simeq \langle f_{q}^{}\rangle \frac{M^{q-1}}{\langle N\rangle^{q}},
\label{DFq-1} 
\end{equation} 
where $\langle N\rangle=\sum n_{i}$ is the average total multiplicity for $M=1$, and $\langle f_{q}\rangle = \langle n(n-1)\ldots(n-q+1) \rangle $ is the factorial moment averaged over the multiplicity distribution in a cell containing a cluster of correlated particles.

In general, for a multiplicity distribution with an average multiplicity $m$, the factorial moment is given by $f_{q} \propto m^{q}$
\footnote{For a multiplicity distribution with a generating function $G(z)$, the $q$-th factorial moment is obtained as $f_q  =d^qG(z)/dz^q \mid_{z=1}$. Specifically, for a Poisson distribution, $f_q =m^q$; for a geometric distribution, $f_q=\Gamma(q+1) m^q$; and for a negative binomial distribution with shape parameter $k$, $f_q =\frac{\Gamma(k+q)}{\Gamma(k)} \left(\frac{m}{k}\right)^q$, where $m$ is the mean multiplicity.}.
Thus, the factorial moment difference follows the relation: 
\begin{equation} 
\Delta F_{q} \propto m^q M^{q-1}. 
\label{DFq-2} 
\end{equation}


\section{Quasi power-law behavior for count-in-cell}
\label{sum-res}

\begin{figure}
\begin{center}
\includegraphics[scale=0.48]{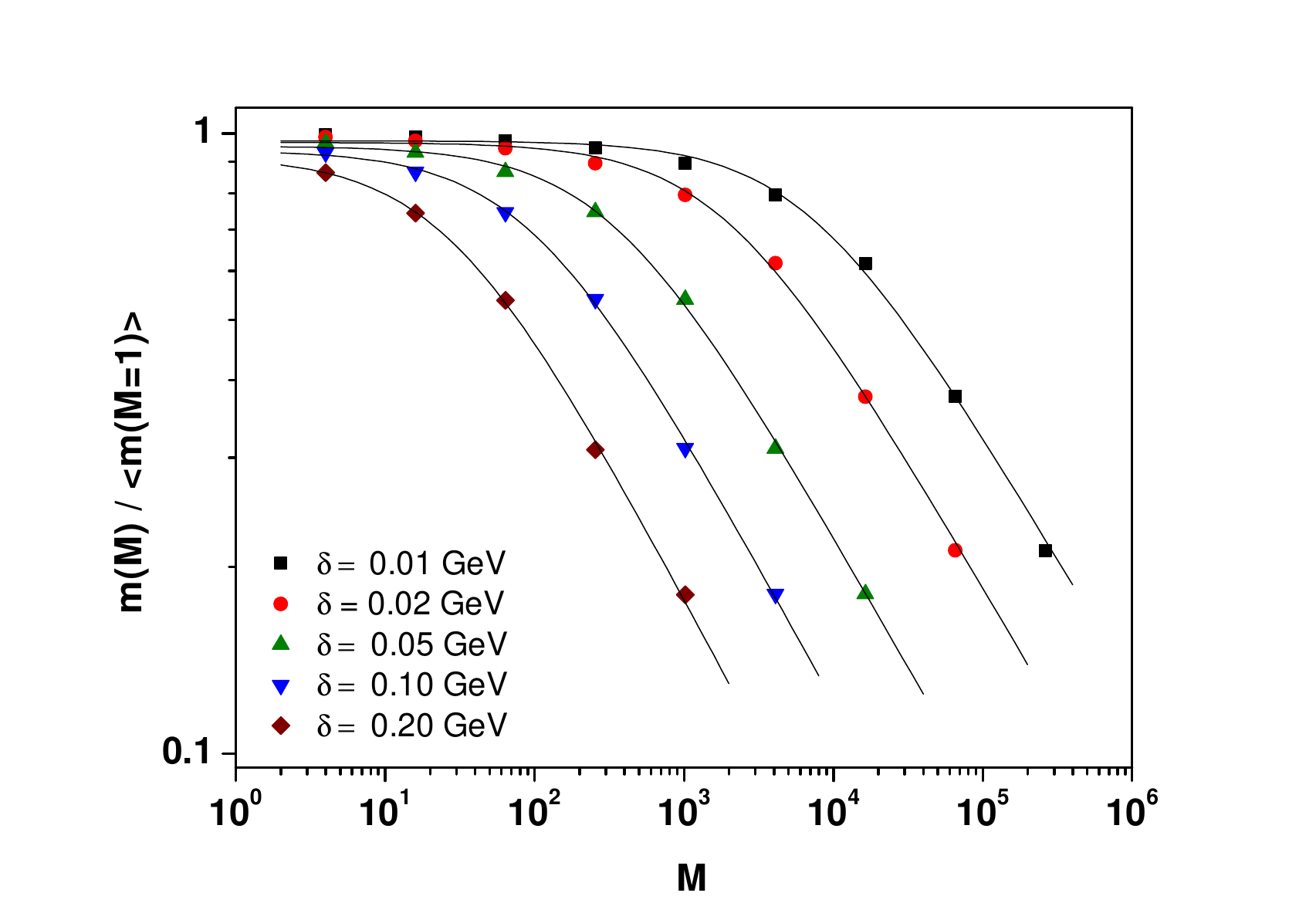}
\end{center}
\vspace{-5mm}
\caption{The dependence of the relative multiplicity within a cluster, $m\left(M\right)/\langle m\left(M=1\right) \rangle$, on the number of cells $M$ is analyzed for different spatial sizes $\delta$. The numerical results are compared with the functional dependence $m(M)$ described by Eq.~(\ref{m}), represented by the lines in the plot.}
\label{SPD-1}
\end{figure}

\begin{figure}
\begin{center}
\includegraphics[scale=0.48]{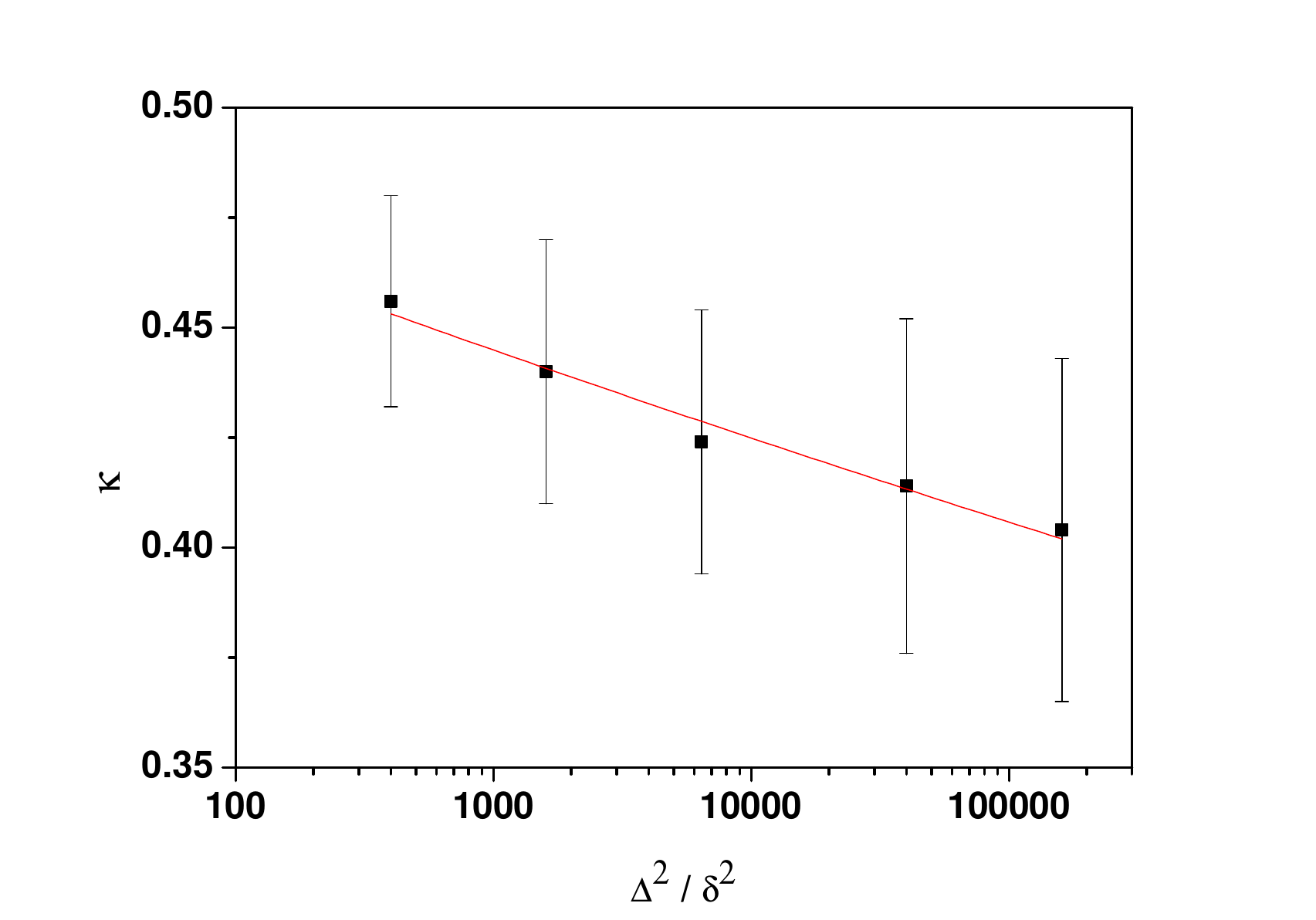}
\end{center}
\vspace{-5mm}
\caption{The dependence of the power index $\kappa$ on the transverse size $\delta$ of the particle cluster is analyzed. The fitted function follows the relation $\kappa = a(\Delta^2/\delta^2)^b$, with $\Delta = 4$ GeV, $a = 0.5$, and $b = -0.02$.}
\label{SPD-2}
\end{figure}
\begin{figure}
\begin{center}
\includegraphics[scale=0.48]{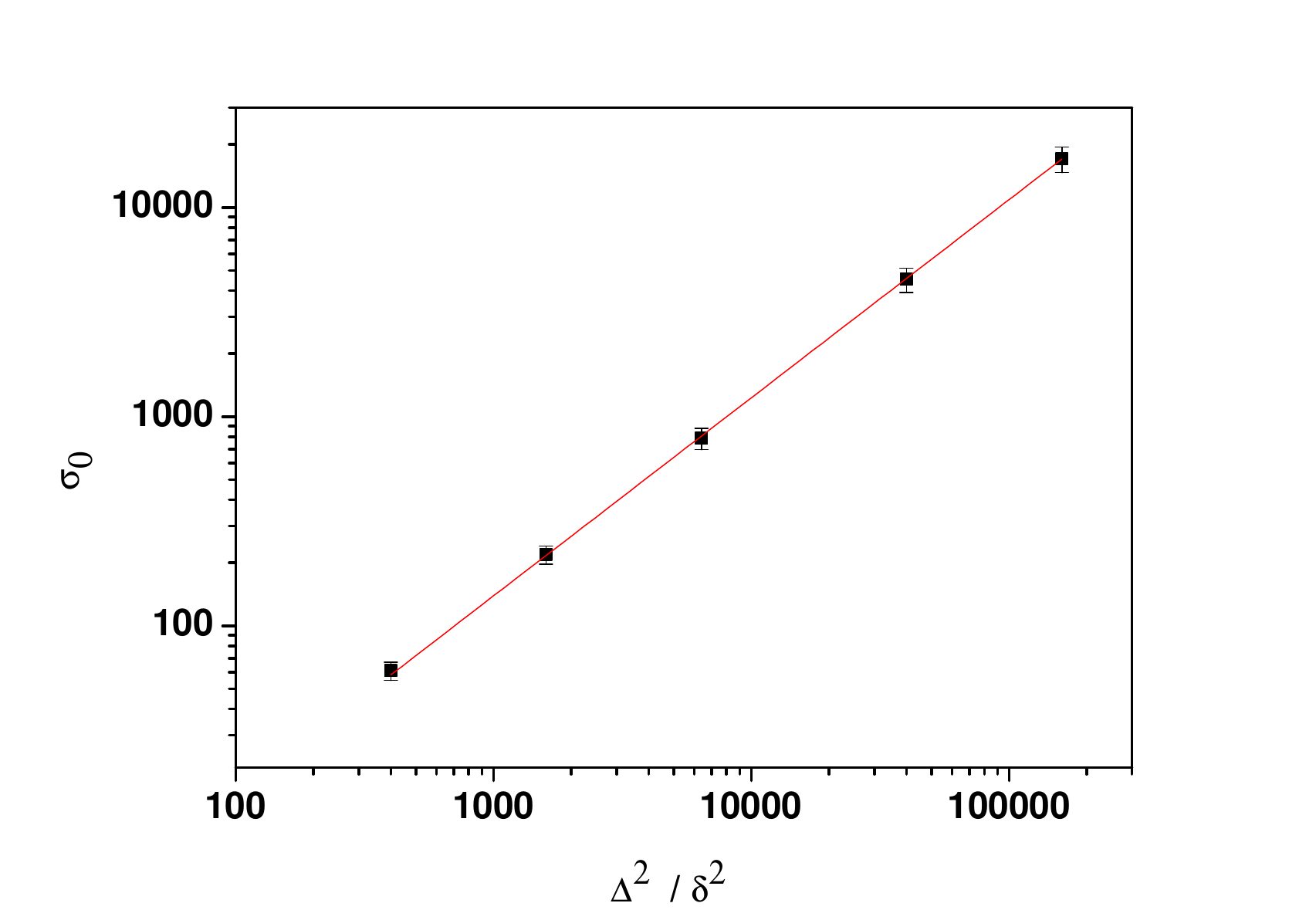}
\end{center}
\vspace{-5mm}
\caption{The dependence of the scale parameter $\sigma_0$ on the transverse size $\delta$ of the particle cluster is examined. The fitted function follows the relation $\sigma_0 = a(\Delta^2/\delta^2)^b$, with $\Delta = 4$ GeV, $a = 0.2$, and $b = 0.947$.}
\label{SPD-3}
\end{figure}

In a scenario where a cluster of particles is confined to a single cell (or a limited number of cells), Eq.~(\ref{DFq-1}) predicts a power-law behavior of the scaled factorial moments, given by $\Delta F_{q} \propto M^{\Phi_{q}}$, 
with the exponent $\Phi_{q} = q-1$, assuming that the multiplicity $m$ of particles in a cluster remains independent of the number of cells $M$. However, in practice, the phase-space division procedure affects the multiplicity $m$ within a given cell. Specifically, the number of particles in a cluster of effective spatial size $\delta$, contained within a phase-space cell of size $\Delta\left(M\right)=\Delta\left(M=1\right)/M$, is expected to vary as $m=m(\delta/\Delta(M))=m(M)$.

The partitioning of particles within a given cluster follows a statistical process governed by a deformed exponential rate. This implies that the number of particles removed from a given cell as $M$ increases is proportional to the number of particles originally present in that cell, leading to the relation:
\begin{equation} 
\frac{dm}{dM}=-\frac{m}{\sigma}. 
\label{dm} 
\end{equation}
In this cell division framework, the division rate $\sigma=\sigma(M)$ depends on the number of cells $M$, since a larger number of cells corresponds to smaller individual cell sizes, increasing the probability that particles within 
a cluster become distributed among multiple cells. A specific form for $\sigma(M)$ has been taken into account:
\begin{equation} 
\sigma\left(M\right)=\sigma_{0}+\frac{M}{\kappa}, 
\label{sigma} 
\end{equation}
where $\sigma_{0}$ is an initial scale parameter, and $\kappa$ controls the deformation of the exponential function. The solution to Eq.~(\ref{dm}) under this assumption yields a $\kappa$-deformed exponential function \footnote{The $\kappa$-deformed exponential function is widelly discussed in Tsallis statistics, where the nonextensivity parameter is equal to $1+\kappa^{-1}$ \cite{Tsallis}.}:
\begin{equation} 
m\left(M\right)\propto \left[1+\frac{M}{\sigma_{0}\kappa}\right]^{-\kappa}, 
\label{m} 
\end{equation}
which reduces to a standard exponential function in the limit $\kappa \rightarrow \infty$.
For a sufficiently large number of cells, $M\gg \sigma_{0}\kappa$, the dependence transitions to a scale-free power law:
\footnote{If the function $m(M)$ follows a pure power law, $m(M) \propto M^{-\kappa}$, it is scale-invariant and satisfies the relation $m(\lambda M) = \mu m(M)$, where the exponent $\kappa$ is given by $\kappa = -\ln \mu /\ln \lambda$~\cite{Sornette}.}
\begin{equation} 
m\left(M\right) \propto M^{- \kappa}. 
\label{mdep} 
\end{equation}

To test this scenario, we performed Monte Carlo simulations. We considered a cluster of particles with a transverse spatial dispersion $\delta$ (relative to the cluster axis) in a randomly chosen direction $(p_{Tx0},p_{Ty0})$, following a Gaussian transverse momentum distribution:
\begin{equation} 
f(p_{Tx},p_{Ty})=\frac{1}{2\pi \delta^{2}} \exp\left(- \frac{(p_{Tx}-p_{Tx0})^{2}+(p_{Ty}-p_{Ty0})^{2}}{2\delta^2} \right). 
\label{pt} 
\end{equation}

The cluster axes were randomly placed within a transverse momentum region of size $\Delta \cdot \Delta$ (where $\Delta = 4$ GeV). For a given mean multiplicity $m_0 = \langle m(M=1) \rangle$, the initial multiplicity $m(M=1)$ was sampled from various multiplicity distributions $P(m)$. 
The numerical results for the dependence of $m(M)$ are presented in Fig.~\ref{SPD-1}. The $\kappa$-deformed exponential function described by Eq.~(\ref{m}), with the exponent $\kappa$ and scale parameter $\sigma_0$  as shown in Figs.~\ref{SPD-2} and~\ref{SPD-3},  provides a successful fit to the numerical results.
For large values of $M$, the expected power-law dependence given by Eq.~(\ref{mdep}) is observed. While variations in the functins $f(p_{Tx},p_{Ty})$ and $P(m)$ affect specific details of the distribution, the underlying scale-invariant behavior remains robust.

\begin{figure}
\begin{center}
\includegraphics[scale=0.48]{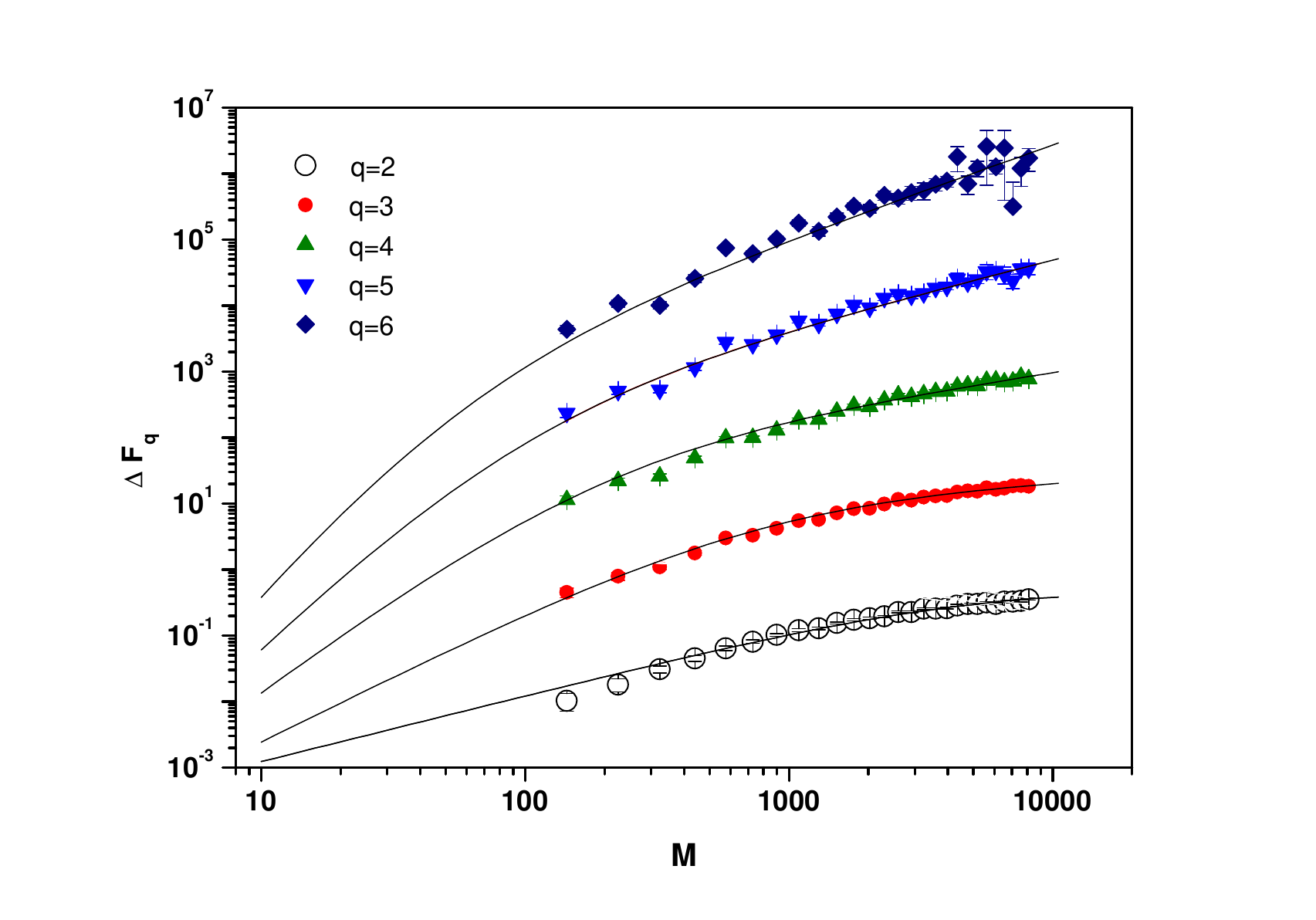}
\end{center}
\vspace{-5mm}
\caption{The scaled factorial moment $\Delta F_{q}$,   as given by Eq.~(\ref{DFq-3}), compared with experimental data from Au+Au collisions at $\sqrt{s_{NN}}=7.7$ GeV~\cite{STAR:2023jpm}.}
\label{SPD-4}
\end{figure}


\begin{table}[]
\centering
\caption{The parameters of Eq.~(\ref{DFq-kappa}) used for the fits presented in Fig.~\ref{SPD-4}.}
\label{tab}
\begin{tabular}{|c|c|c|c|c|}
\hline
\boldmath{$q$} & \boldmath{$C_q$} & \boldmath{$\sigma_0$} & \boldmath{$\kappa$} & \boldmath{$C_q (\sigma_0 \kappa)^{q \kappa}$} \\ \hline \hline
$2$ & $1.2\cdot 10^{-4}$ & $10000$ & $0.6$ & $4.10$ \\ \hline
$3$ & $3\cdot 10^{-5}$   & $1226$  & $0.6$ & $4.34$ \\ \hline
$4$ & $2\cdot 10^{-5}$   & $302$   & $0.6$ & $5.26$ \\ \hline
$5$ & $8.5\cdot 10^{-6}$ & $140$   & $0.6$ & $5.04$ \\ \hline
$6$ & $7.79\cdot 10^{-6}$& $75$    & $0.6$ & $6.96$ \\ \hline
\end{tabular}
\end{table}

\begin{figure}
\begin{center}
\includegraphics[scale=0.48]{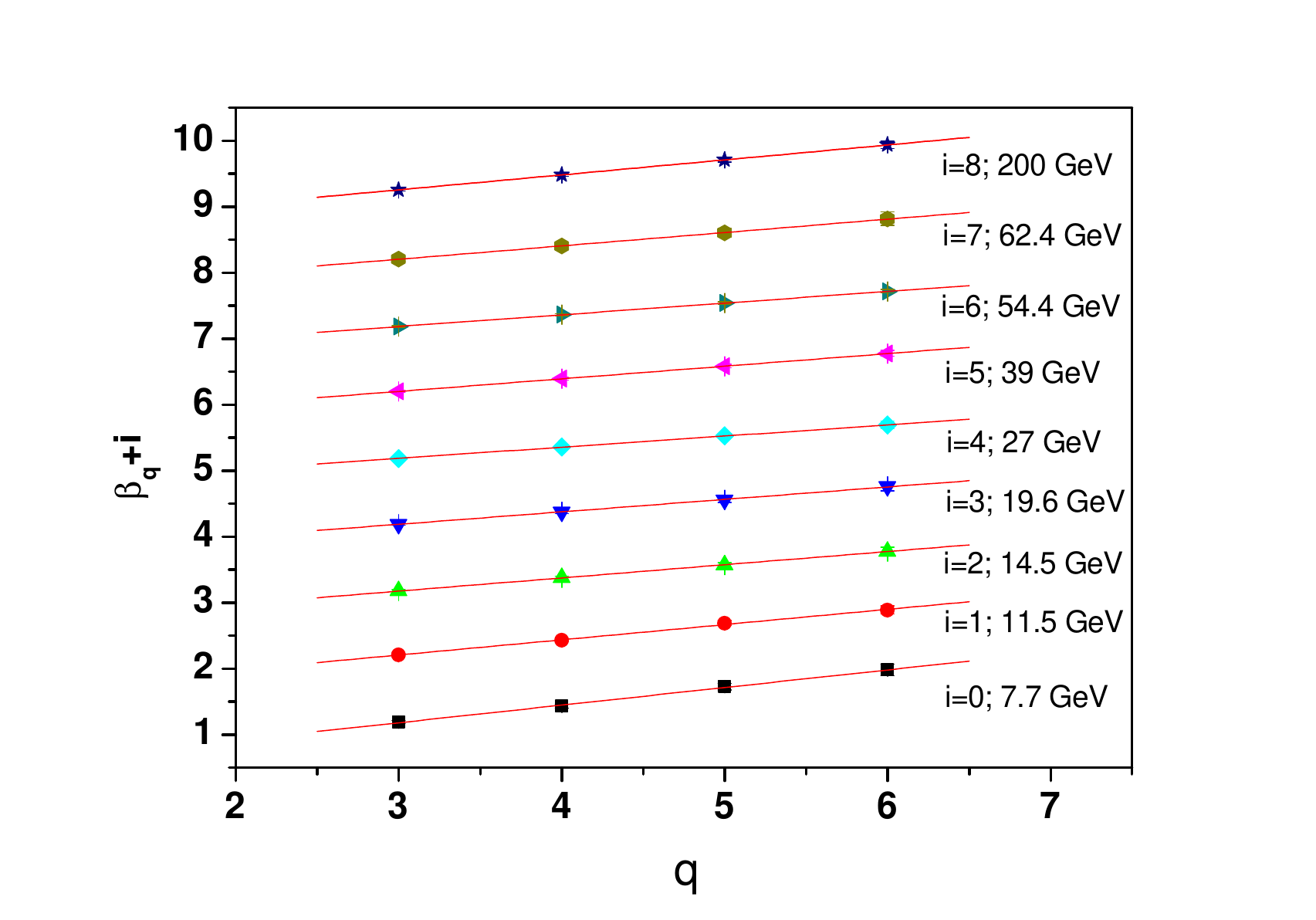}
\end{center}
\vspace{-5mm}
\caption{The scaling index $\beta_q$ ($q=3$–$6$) is shown as a function of $q$ for the most central ($0$–$5\%$) Au+Au collisions at $\sqrt{s_{NN}} = 7.7$–$200$ GeV~\cite{STAR:2023jpm}. For improved readability, the data points corresponding to different energies have been shifted by $i=0,1,\dots,8$, respectively. The lines represent the linear fit described by $\beta_q = a(q-2) + b$.} 
\label{SPD-6}
\end{figure}
\begin{figure}
\begin{center}
\includegraphics[scale=0.48]{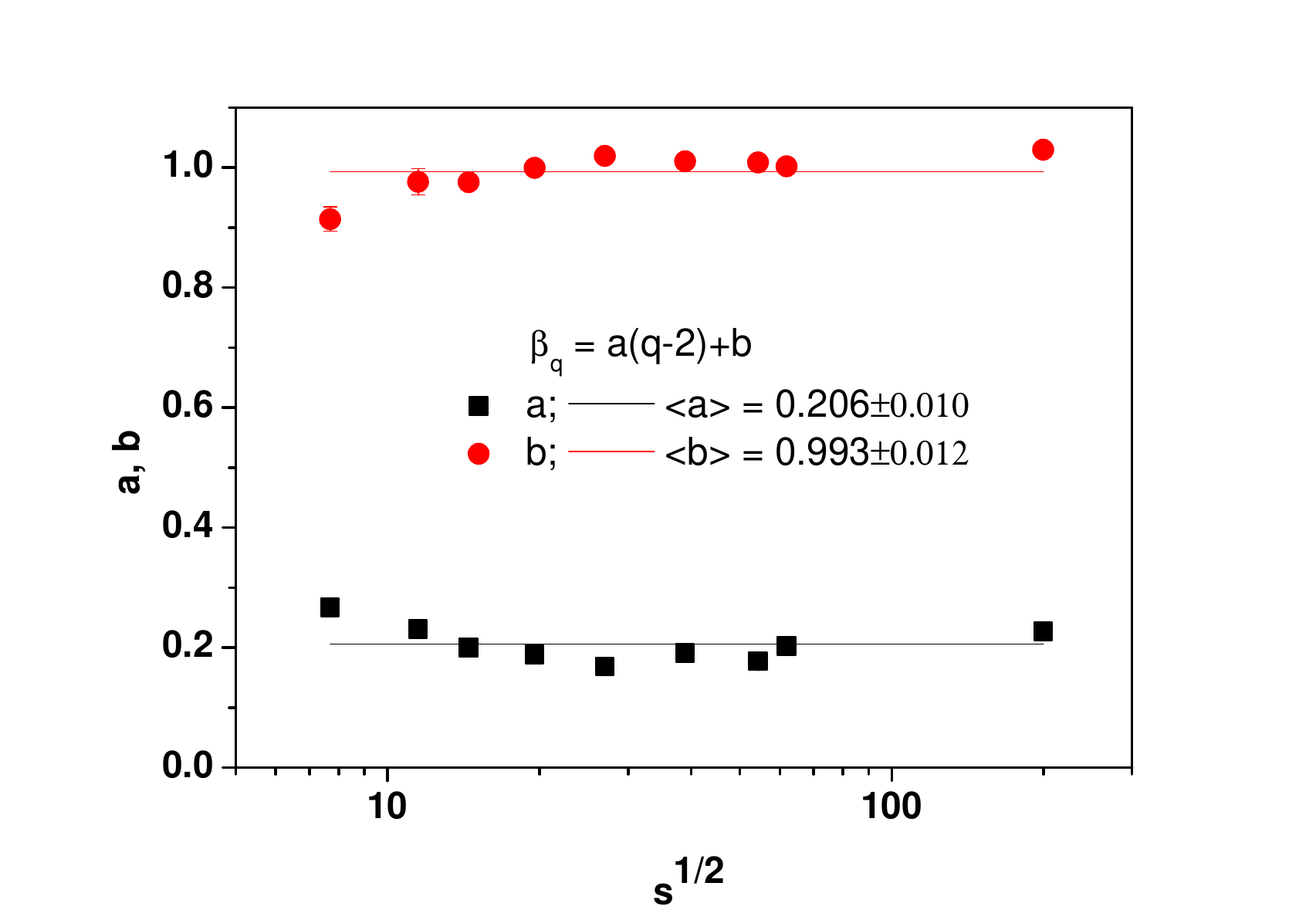}
\end{center}
\vspace{-5mm}
\caption{The parameters of the linear relation $\beta_q = a(q-2) + b$, shown in Fig.~\ref{SPD-6}, are presented as a function of the collision energy $\sqrt{s_{NN}}$.}
\label{SPD-7}
\end{figure}

\section{Comparison with intermittency results}
\label{comparison}

The scaled factorial moments given by Eq.~(\ref{DFq-2}), with the multiplicity $m(M)$ described by Eq.~(\ref{m}), result in the following expression:
\begin{equation} 
\Delta F_{q}\left(M\right) = C_q m^q(M) M^{q-1} = C_q \left [1+\frac{M}{\sigma_0 \kappa} \right ]^{-q \kappa} M^{q-1}, 
\label{DFq-kappa} 
\end{equation}
where $C_q \propto m_0^q\langle N \rangle^{-q}$ is a normalization factor that does not depend on the number of cells.
For large $M$, the dependence transitions to a power-law form:
\begin{equation} 
\Delta F_{q}\left(M\right) = C_q (\sigma_0 \kappa)^{q \kappa} M^{q(1-\kappa)-1}. 
\label{DFq-3} 
\end{equation}
The exponent $\Phi_{q} = q(1-\kappa)-1$ arises as a direct consequence of the phase-space division procedure, which leads to the power-law dependence of $m(M)$ given in Eq.~(\ref{mdep}).

The STAR Collaboration reported extensive measurements of $\Delta F_{q}\left(M\right)$ in the two-dimensional transverse momentum plane $(p_{Tx}, p_{Ty})$ for central Au+Au collisions at center-of-mass energies $\sqrt{s_{NN}} = 7.7$, 11.5, 19.6, 27, 39, 62.4, and 200 GeV~\cite{STAR:2023jpm}.

In Fig.~\ref{SPD-4}, we compare the experimental data with the dependence predicted by Eq.~(\ref{DFq-kappa}). The fitted parameters are presented in Table~\ref{tab}. Independent on the number of cells, the prefactor $C_q (\sigma_0 \kappa)^{q \kappa} \approx 2.5 + 0.66 q$ exhibits only a weak dependence on $q$. Additional experimental data at lower values of $M$ (particularly for $M < 100$) would be valuable for refining the parameter estimates.

For the power-law dependence given in Eq.~(\ref{DFq-3}), the ratio of scaled factorial moments satisfies the relation:
\begin{equation} 
\frac{\Delta F_{q}}{\Delta F_{q'}} \propto M^{(q-q')(1-\kappa)}, 
\label{ratio} 
\end{equation}
which implies the scaling behavior described in Eq.~(\ref{Fq3}), where the exponent $\beta_q$ exhibits a linear dependence on the moment order $q$. For values of $q \leq 6$, a linear function of the form $\beta_q = a(q-2) + b$ provides a good fit to the experimental data. In Fig.~\ref{SPD-6}, we present a fit of the experimental $\beta_q$ values using this functional form, with the corresponding fit parameters displayed in Fig.~\ref{SPD-7}. Approximate scaling behavior is observed, with $\beta_q$ well described by the relation $\beta_q \simeq 0.2(q-2) + 1$, for collision energies in the range $\sqrt{s_{NN}} = 7.7 - 200$ GeV.

\section{Conclusions}
\label{Concl}

The phase-space division procedure leads to a power-law-like behavior of scaled factorial moments. A correlated cluster of particles confined within a single cell can give rise to an "intermittency-like" pattern in $\Delta F_{q}\left(M\right)$. For sufficiently small cell sizes, $\Delta^2/M$, the multiplicity within a cluster of size $\delta$ exhibits scale-invariant behavior, following a simple power law, $m(M) \propto M^{-\kappa}$. Consequently, the scaled factorial moments obey the relation $\Delta F_{q}\left(M \right) \propto M^{q(1-\kappa)-1}$.

We demonstrate that the observed power-law behavior of scaled factorial moments is not exclusive to critical phenomena. To determine whether this simplified scenario can account for the broad range of experimental observations, further calculations based on dynamical modeling of heavy-ion collisions are necessary.

\vspace*{0.3cm}
\centerline{\bf Acknowledgments}
\vspace*{0.3cm}
This research was supported by the Polish National Science Centre (NCN) Grant 2020/39/O/ST2/00277. In preparation of this publication we used the resources of the Centre for Computation and Computer Modelling of the Faculty of Exact and Natural Sciences of the Jan Kochanowski University in Kielce, modernised from the funds of the Polish Ministry of Science and Higher Education in the “Regional Excellence Initiative” programme under the project RID/SP/00015/2024/01.




\end{document}